# SpRRAM: A Predefined Sparsity Based Memristive Neuromorphic Circuit for Low Power Application

Arash Fayyazi*     Souvik Kundu*     Shahin Nazarian     Peter A. Beerel     Massoud Pedram

University of Southern California, Los Angeles, USA

{fayyazi, souvikku, s.nazarian, pabeerel, pedram}@usc.edu

*Abstract—* In this paper, we propose an efficient predefined structured sparsity-based ex-situ training framework for a hybrid CMOS-memristive neuromorphic hardware for deep neural network to significantly lower the power consumption and computational complexity and improve scalability. The structure is verified on a wide range of datasets including MNIST handwritten recognition, breast cancer prediction, and mobile health monitoring. The results of this study show that compared to its fully connected version, the proposed structure provides significant power reduction while maintaining high classification accuracy.

## I. INTRODUCTION

In today's data-driven world, Deep Neural Networks (DNNs) play a key role in driving the state of the art technologies like image processing, pattern recognition and speech recognition. Modern neural networks are formally built as graphs with a large number of trainable parameters [1] which is memory intensive. This makes the process of both training and inference of such networks an arduous job to perform in a power efficient on-chip manner.

Imitating the computational behaviors and reverse engineering the immensely authoritative cognitive capacities of the human brain has opened the door of brain-inspired neuromorphic computing [2] which has two major variant as Artificial Neural Network (ANN) and Spiking Neural Network (SNN) [3]. In our paper, we have focused on analog implementation of ANN for its higher accuracy and power efficiency [3]. Memristor [4], also known as the fourth fundamental passive two-terminal component is one of the most popular choices for analog ANN design due to its area efficient way of storing weights in terms of resistance.

Introduced in 1971 and first physical realized in 2008 [5], this element has the capability of being fabricated densely and has plasticity in resistance [6]. The conductance value of the memristors can be altered by the voltage applied to them that should be larger than a threshold voltage. The amplitude and duration of the voltage pulse determine the amount of the change in the resistance. We have used a cross-bar array of memristors for hardware realization of synapses of ANNs which performs a weighted summation of the inputs. In addition to the synapses, an activation function which is usually a non-linear function is essential for implementation of an ANN. The activation function can be implemented using circuits such as CMOS inverters [7] and op-amps [8]. In this work, we have focused on the inverter-based implementation which would result in lower area and power consumption [7].

The large number of parameters and connectivity of the network provide the neural network a complex non-linear nature that enables them to effectively draw precise decision boundary between classes; however, too many parameters make the network likely to overfit the outputs of the ANN to the training samples. This tradeoff makes it a challenging task to predict a proper network structure for a dataset. For many overfitted networks separate hyperparameters are used to reduce overfitting and remove unnecessary memorization of undesirable noise patterns. Earlier methods of parameter reduction mentioned in [9],[10] perform significantly well in software but are not quite helpful to adapt in a hardware implementation of network training because they require the network to be fully connected at some point in time during training. Also, their additional computation overhead to reduce network structure causes extra power consumption which limits their applicability in a power-sensitive hardware environment. Recently, [11] has shown that many of these parameters can be omitted without any significant loss in network fidelity. In this paper, we adopt the idea of a hardware implementable predefined sparsity (keeping a fraction of the total weights, in one or multiple junctions before training) [12] based neural network training. In particular, we propose a power efficient structured predefined sparsity-based hybrid memristive ex-situ training framework. Here, by 'structured' we mean equal fan-out for each neuron in the preceding layer and equal fan-in for each neuron in succeeding layer respectively. This framework is unique because the training algorithm can also be implemented on-chip without any additional computation or hardware cost to reduce the number of weights. In our proposed framework, the Memristor Crossbar Array (MCA) is no longer densely connected. In addition to being power efficient, these sparsely connected MCAs inherently deal with the overfitting issue of a DNN.

The major contributions of this paper are summarized below:

1. To the best of our knowledge, we are the first to propose and validate a structured predefined sparse network in a MCA-based ex-situ training framework. This approach significantly reduces the power consumption of large neural networks for on-chip inference, making it a strong hardware foundation for ultra-low power machine learning domains, such as in IoT edge devices.
2. We present an enhancement to the Physical Characteristic Aware Ex-situ training framework (PHAX) [7] to support sparse ANN structures. Using this framework, we show that predefined sparsity can yield no significant loss of test accuracy and cross-validated these results with a HSPICE simulation model.
3. We further present experimental results that explore the effect of different levels of sparsity and memristor process variation on classification accuracy of the network.

---
* Equal contribution

The rest of the paper is organized as follows. Section II presents the proposed predefined sparsity-based ex-situ memristive training framework. Section III describes the experimental setup and results against benchmark datasets (IRIS [13], BCW [14], MNIST [15] and MHEALTH [16]). Finally, the paper is concluded in section IV.

## II. PROPOSED FRAMEWORK

We first review the MCA present in PHAX [7] in Section II.A before describing our enhancement to support predefined sparsity in inference and training in Section II.B and II.C respectively. It is noteworthy that our proposed training algorithm is applicable to a variety of underlying memristive circuits [7], [8].

### A. Memristor-based ANN Structure

The fully connected (FC) junction memristive circuit, shown in Fig. 1, has a better performance, lower power consumption, higher energy efficiency, and smaller area than other memristive neuromorphic circuits such as the op-amp based circuit proposed in [8]. For this circuit, the dot-product operation is performed in the memristive crossbar while the inverters implement the neuron's nonlinear operation, i.e. activation function. The circuit, which has differential inputs, makes use of two memristors per weight, implementing both negative $(n)$ and positive $(p)$ weights.

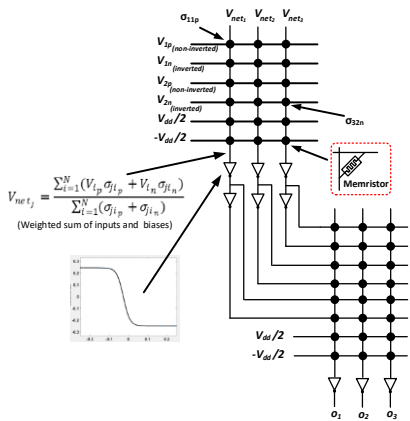

Fig. 1. Circuit structure of the FC memristive neuromorphic circuit [7] used in this work.

The inverter voltage transfer characteristics (VTC) has the form of a scaled sigmoidal function, which acts as the neuron activation function. To provide differential inputs for the next layer, two inverters are used at the output of each layer (except the last layer). It should be noted that for classifier ANNs, each output of the circuit corresponds to one of the output classes and has a digital value of either logical "0" or logical "1". In the case of function approximate applications, however, the outputs of ANNs are analog and an analog-to-digital converter is needed to digitalize their outputs [17].

### B. Structured Predefined Sparsity Characterization

A DNN with all FC junctions has all neurons of the $(l-1)^{th}$ layer connected to all neurons of the $(l)^{th}$ layer, thus demanding a large memory to store the weights of each junction which must be updated during the training. Most parameter reduction methods such as [9],[10] reduce the size of these weight matrices through iterative algorithms and start with an FC network. Hence, it is not clear how to apply these methods to a hardware training environment. To address this issue, we propose a structured predefined sparse neural network. Consider a network with $L$ layers of neurons. Thus, the network has $J = (L-1)$ junctions. The junction between layer $j$ and layer $j+1$ has $N_j \times N_{j+1}$ weights giving a total of $\sum_j N_j \times N_{j+1}$ for an FC network where $N_j$ is the number of neurons of layer $j$. We take the connection density of a junction as defined in [12],

$$D_j = (W_j / N_j \times N_{j+1}) \tag{1}$$

where $W_j = N_j \times FO_j = N_{j+1} \times FI_j$, and $FO_j$ and $FI_j$ are fan-out count and fan-in count from each of preceding and succeeding neurons, respectively.

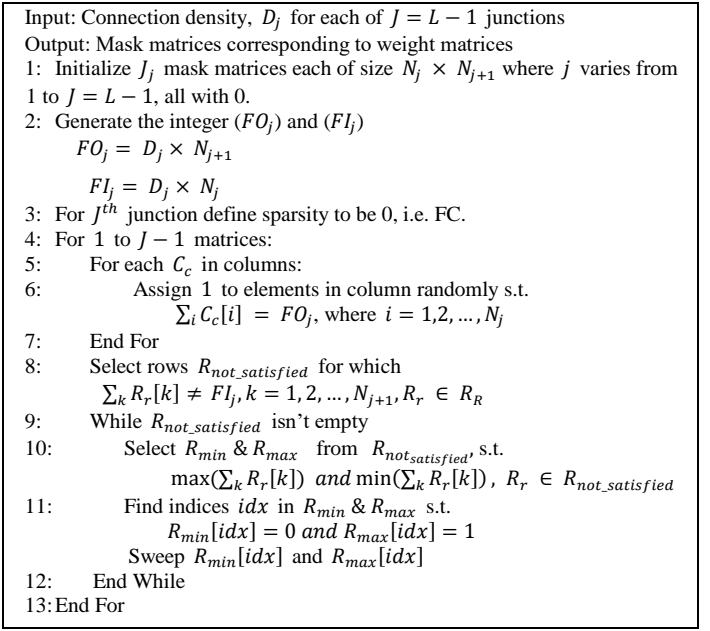

Fig. 2 Pseudocode for generating the mask matrices.

For predefined sparsity in our work, we arbitrarily remove the edges between two successive layers by following algorithm shown in Fig. 2 and generate the mask matrices

(matrices which take care of which weights to keep) and then train the network with remaining connections (weights) using the generated mask matrices. Therefore, for this kind of network reduction technique, the removed weights never appear during training or inference and hence this technique is a preferable choice for hardware design.

Fig. 3 explains the difference between an FC junction, an unstructured sparse junction and a structured sparse junction with 50% connectivity (i.e. 50% sparsity). It is to be noted that we have kept the bias connection out of the scope of sparsity, i.e. the biases for all the neurons are always present.

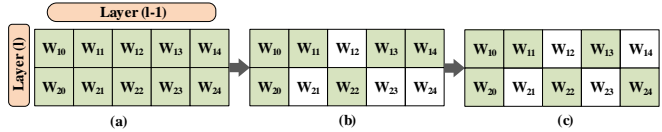

Fig. 3. Generation of a predefined sparse structure of a network with four neurons in layer $l-1$ and two in layer $l$, (a) a fully connected (FC) junction between layer $l-1$ and $l$, (b) its unstructured sparsity version with a predefined connectivity of 50%, (c) corresponding structured sparse version.

The structured sparse version can be used as a mapping scheme that can map a given DNN to any MCA size permissible by the memristive technology for reliable operations. As shown in Fig. 4, we have a network with a junction of 4:2 (4 neurons in layer $l-1$ and 2 neurons in layer $l$), with proposed structured sparse version, we can easily map it to two 2:1 MCA. Thus, this mapping technique has a potential advantage of saving area.

To consider the predefined sparse connections, we have removed both the memristors per weight according to mask matrices. Fig. 5 show a structured sparsity version of the memristive circuit with a predefined connectivity of 25%. Each of the $N$ inputs of this circuit (e.g., $x_i$) is differential containing inverted ($V_{i_n}$) and non-inverted ($V_{i_p}$) signals. The Kirchhoff's current law (KCL) at the input of the $j^{th}$ inverter may be written as (see Fig. 5)

$$\sum_{i=1}^{N} a_{ji}\left((V_{i_p} - V_{net_j})\sigma_{ji_p} + (V_{i_n} - V_{net_j})\sigma_{ji_n}\right) = 0 \quad (2)$$

Where $a_{ji}$ has a value of either 0 or 1 to indicate whether there is a connection between the $j^{th}$ neuron of the current layer and $i^{th}$ neuron of the previous layer. Also, $V_{net_j}$ is the voltage of the node $net_j$ (the input of the inverter of column $j$), and $\sigma_{ji_p}$ ($\sigma_{ji_n}$) is the conductance of the memristor located in the non-inverted (inverted) row $i$ and column $j$. Therefore, the input voltage of the $j^{th}$ inverter ($V_{net_j}$) can be obtained from

$$V_{net_j} = \frac{\sum_{i=1}^{N} a_{ji}(V_{i_p}\sigma_{ji_p} + V_{i_n}\sigma_{ji_n})}{\sum_{i=1}^{N} a_{ji}(\sigma_{ji_p} + \sigma_{ji_n})} \quad (3)$$

Also, assuming a neural network layer with the inputs $V_{i_p}$ and $V_{i_n}$ ($i=1, \ldots, N$) and the weights of $w_{ji_p}$ and $w_{ji_n}$ weights, the input of the $j^{th}$ neuron ($net_j$) of this layer may be obtained from

$$net_j = \sum_{i=1}^{N} a_{ji}(V_{i_p}w_{ji_p} + V_{i_n}w_{ji_n}) \quad (4)$$

Note that in (2)-(4), the positive (negative) bias is considered as an input with a constant value of $V_{dd}/2$ ($-V_{dd}/2$).

For the crossbar circuit with the inverter-based neurons to mimic the neural network, the values of $net_j$ and $V_{net_j}$ should be the same for all the combinations of the inputs. Therefore, the corresponding coefficients (i.e., $w_{ji_p}$ and $w_{ji_n}$) are expressed as

$$\begin{cases} w_{ji_p} = \frac{\sigma_{ji_p}}{\sum_{m=1}^{N} a_{jm}(\sigma_{jm_p} + \sigma_{jm_n})}, & a_{ji} == 1 \\ w_{ji_p} = 0, & a_{ji} == 0 \end{cases} \quad (5)$$

Thus,

$$a_{ji}\left[w_{ji_p}\left(\sum_{m=1}^{N} a_{jm}(\sigma_{jm_p} + \sigma_{jm_n})\right) - \sigma_{ji_p}\right] = 0 \quad (6)$$

We may find the relations between weights and memristor conductance by solving equations in (6) which can be shown that has a non-trivial solution only if the

$$\sum_{i=1}^{N} a_{ji}(w_{ji_n} + w_{ji_p}) = 1 \quad (7)$$

Hence, the learning algorithm should train the neural network with these constraints:
1. Weights must be positive and the conductance of all of the memristors must be in the range of $[\sigma_{min}, \sigma_{max}]$. Where $\sigma_{min}$ and $\sigma_{max}$ are minimum and maximum of possible conductance values. In this work, we use the memristor parameters of [18] where the maximum and minimum values are 7.9μ℧ and 0.12μ℧, respectively.
2. The sum of all the existing weights must be equal to 1 (based on (7)).

C. *Training of Network with Predefined Sparsity*

To train the neural network, we propose an ex-situ training algorithm based on PHAX [7]. PHAX [7] is a circuit aware training framework where the backpropagation algorithm of [19] was modified to consider the physical characteristics of the neuromorphic circuit. Our proposed algorithm benefits from the integrity of the gradient descent search in the backpropagation algorithm as well as effectively dealing with the physical constraints (described in Section II.B) in the memristive circuits. The pseudocode for the proposed algorithm is given in Fig. 6. In our proposed algorithm, we check the existence of connection in forward and backward. Since the proposed weight mapping function has components of other weights in the same layer (in other words, all the neurons in the previous layer), the weights of the removed connection must keep zero before and after applying weight mapping functions (line 3 and 13 in pseudo code.)

In the proposed algorithm, $\theta_{kj}$ is the weight between the $j^{th}$ neuron of the previous layer $(l-1)$ and the $k^{th}$ neuron of the current layer $(l)$, $g_1$ and $g_2$ are two mapping functions to consider the physical constraints[7] and maps unconstrained ANN weights to implementation weights (memristor conductance in this work) and are incorporated in the update rules of the backpropagation algorithm. Also, $\delta_k$ is the portion of the error of the $k^{th}$ neuron of the output layer, $\eta$ is the learning rate and $O_m$ and $t_m$ are the actual and target output of the $m^{th}$ neuron. Finally, $net_m$ is the weighted sum of the inputs of the $m^{th}$ neuron. Although the above equations have been expressed for the case of the $P$ weights, very similar equations may be written for the $N$ weights. Note that we used the same notation as [7] where PHAX algorithm is discussed in deep. In this work, we set the target accuracy to 98% (classification error of 2%) and epoch limit to 100,000.

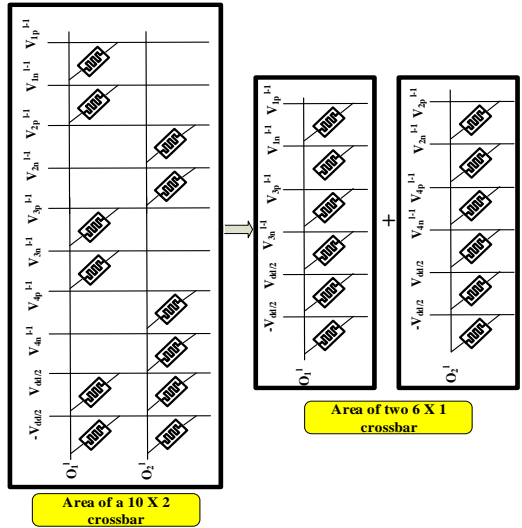

Fig. 4. Structured sparse version with 50% of connection and potential clustering by utilizing structured sparse version and smaller MCA (2 of 2:1.)

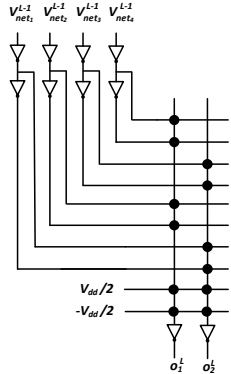

Fig. 5. Circuit structure of the memristive neuromorphic circuit with a structured sparsity of 25%. Mask matrix for this layer is $\begin{bmatrix} 1 & 0 \\ 0 & 1 \\ 1 & 0 \\ 0 & 1 \end{bmatrix}$

## III. EXPERIMENTAL RESULTS

In this section, we present the performance of our proposed framework in terms of inference accuracy in benchmark applications, complexity analysis and power consumption. We have also investigated how accuracy varies w.r.t different percentages of sparsity and sparsity at different junctions and have also analyzed the network performance subject to variation in memristor conductance as well as bit precision limits. We have generated the conductance values from MATLAB and used them in corresponding HSPICE simulation models assuming TSMC 90nm technology and the memristor model proposed in [20] for the memristors devices of [18]. We used SPICE simulation to measure the inference accuracy and power consumption.

```
1:  Initialize all weights with small random numbers
2:  For each layer in the network
3:      θʲ = θʲ ⊙ Aʲ
4:  End For
5:  Do
6:      For every pattern in the training set
7:          Present the pattern to the network
8:          // Propagated the input forward through the network:
9:          For each layer in the network
10:             If a_{ji} == 1 then
11:                 Calculate the function mapped weights w_{ji_p} and w_{ji_n}
                    w_{ji_p} = g_2(σ_{ji_p}) = g_2(g_1(θ_{ji_p}))
12:             Else
13:                 w_{ji_p} = 0
14:             End If
15:             For every node in the layer
16:                 Calculate the weighted sum of the inputs to the node
17:                 Calculate the activation for the node
18:             End For
19:         End For
20:         Calculate the Cost function J
21:         // Propagate the errors backward through the network:
22:         For every node in the output layer
23:             Calculate the error signal (t_k − O_k^L)
24:             If a_{ji} == 1 then
25:                 Calculate the weight updating rule (change value) Δθ_{kj}
26:                 Update each node's weight in the output layer
27:             End If
28:         End For
29:         For all hidden layers
30:             For every node in the layer
31:                 Calculate the node's signal error (∑_{k∈(l+1)} δ_k aw_{ji_p})
32:                 If a_{ji} == 1 then
33:                     Calculate the weight updating rule Δθ_{kj}
34:                     Update each node's weight in the network
                        θ_{kj new} = θ_{kj old} − η Δθ_{kj}
35:                 End If
36:             End For
37:         End For
38:     End For
39: Till (maximum number of iterations > than specified) or
40:     (Cost function J is < than specified))
```

Fig. 6. Pseudocode of the proposed training algorithm.

Our MATLAB script takes user inputs of the level of predefined sparsity and generates mask matrices to use along with the mathematical model of the neurons (*i.e.*, fitted VTC of the inverters that are extracted via SPICE simulation). The script then models the network training and maps the converged trained values to the desired memristor conductance. For the memristor device used in this work, the write threshold voltage was 4V and the minimum (maximum) resistance of the memristors was about 125KΩ (8.3MΩ). In all the simulations, the supply voltage level was 0.5V. Also, the computer system used for the simulations utilized an Intel Core™ i7-7700HQ CPU with a nominal clock frequency of 2.8GHz and 16 GB of RAM. Note that we assumed scheme of [21] to be implemented for writing the memristors which addresses the problems of device variation and stochastic write and has the relative accuracy of 99%. Also, the utilized memristor model has a good endurance and retention [18]. Since the *I-V* characteristic (and hence the conductance) of the memristor is determined by its state variables, the extracted conductance values, the methodology proposed in [7] is used to map the extracted conductance values to the corresponding state parameter of the memristor for the use in the SPICE simulations.

### A. Classification Accuracy with Predefined Sparsity

We have used IRIS, BCW, MNIST and MHEALTH datasets for the performance measurement. 80% data of a dataset are randomly chosen for training whereas the rest 20% are used for test accuracy measurement. Table I provides the details of the network structures for each of the classification dataset and Fig. 7 provides their accuracy under a different percentage of network connections (or in other words, under a different level of sparsity). It is clear from the results in Fig. 7 that with sparse connectivity at the penultimate junction (junction $J-1$ for a network with $J$ junctions) having connectivity as low as 25% the network's classification accuracy is hardly degraded. To train the networks we ran 10,000 epochs and used a mini-batch size of 1.

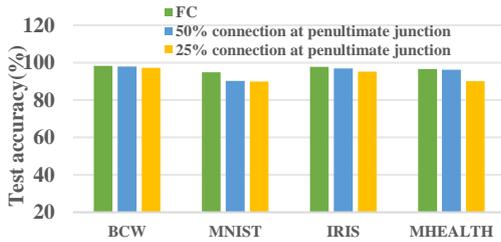

Fig. 7: Relation between test accuracy and the percentage of sparsity in the penultimate junction for various datasets in FC network and sparse network with sparsity applied at penultimate junction only.

For the IRIS, BCW and MNIST datasets we did not observe any considerable difference in the accuracy by exchanging the sparse junction and FC junction between junction 1 & 2, for MHEALTH we did apply different % of sparse connections at junction 2. Another important trend we noticed is for MNIST where applying 50% connectivity at both junctions performs poorly compared to a 25% connectivity followed by an FC junction (or FC followed by 25% connectivity junction).

### B. Power Efficiency and Computation Complexity Reduction

Power and computation complexity in term of memristor counts (memory elements) of the sparse memristive neuromorphic circuit with those of the FC memristive circuit are compared in Fig. 8. Table I mentions the power where the sparse version has 25% connectivity at the penultimate junction. The comparison confirms that using the proposed sparse circuit results in considerable power saving over the fully-connected memristive circuit. The power consumption of the proposed circuit with 25% connectivity in the penultimate junction is reduced by 57% for MNIST dataset.

TABLE I
Network Structures Used for Different Datasets

| Dataset | Network structure | Power Consumption for FC version (µW) | Power consumption for sparse version (µW) |
|---|---|---|---|
| BCW | 10-8-2 | 13.4 | 8.87 |
| MNIST | 196*-100-10 | 1221 | 527 |
| IRIS | 4-4-3 | 7.67 | 7.45 |
| MHEALTH | 23-80-60-13 | 703.2 | 639 |

* A compressed version (14 × 14) of actual 28 × 28 input image is used

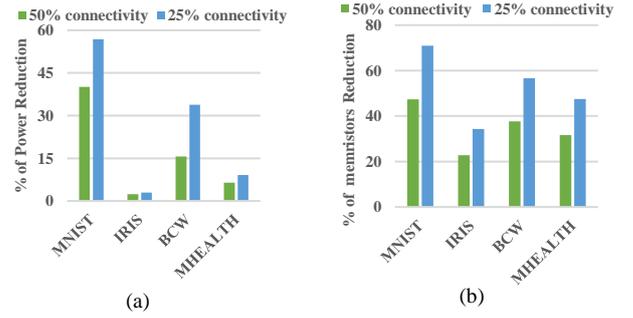

Fig. 8. (a) Power and (b) complexity reduction through reduction of memristor count for proposed sparse neuromorphic circuits over their FC counterparts for different applications. Note that 25% and 50% connectivity is only at the penultimate junction.

Additionally, compared to the FC designs, our proposed sparse design has very low computational complexity in hardware. Note that inverters make up a large portion of the total power consumption.

### C. Effect of Process Variation and Limited Write Precision

To evaluate the performance of the proposed design subject to process variation and limited write precision, we have added random Gaussian noise with 5%, 10%, and 25% variances to the conductance values of the memristors and measured accuracy. For the sake of space, In Fig. 9 we have shown the effects on memristive circuit for only IRIS. With worst-case simulation of 25% process variation results show that the accuracy of the pre-defined sparse structure with only 25% connections in the penultimate junction, is still above 84%, 82.2%, 92%, 95% for MNIST, MHEALTH, IRIS and BCW respectively. Authors of [22], [23] have proposed algorithms to deal with this variability of memristors and CMOS components in an efficient way. [23] uses the neurons characteristics (CMOS inverter here) extracted from a manufactured chip and based on that adjusts weights of the chip.

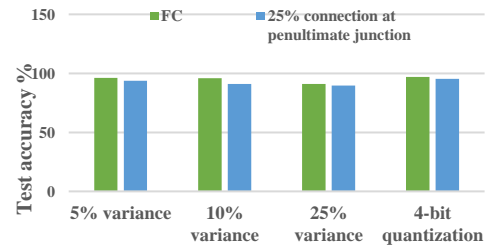

Fig. 9 IRIS dataset classification accuracy in inverter-based FC and sparse memristive neuromorphic circuits under process variations and 4-bit quantization.

IV. SUMMARY AND CONCLUSIONS

In this work, we analyze structured predefined sparse memristor crossbar array structure in ex-situ training framework and its impact on classification accuracy, power, memristors count and process variations of memristor. We obtain a test accuracy of ~90% in MNIST dataset with only ~1/4 of the total weights present. Also, we have obtained a considerable power efficiency of this structure compared to FC memristive neuromorphic circuits. An efficient area optimization technique for the reduced memristor count design is a promising future scope for our work.